\documentclass[twocolumn,preprintnumbers,amsmath,amssymb]{revtex4}
\usepackage{graphicx}
\usepackage{dcolumn}
\usepackage{bm}
\usepackage{times}
\usepackage[colorlinks,citecolor=blue,linkcolor=red]{hyperref}
\usepackage{color}
\usepackage{comment}

\begin{document}

\title{Managing quantum heat transfer in nonequilibrium qubit-phonon hybrid system}

\author{Chen Wang$^{1,}$}\email{wangchenyifang@gmail.com}
\author{Lu-Qin Wang$^{2}$}
\author{Jie Ren$^{2,}$}\email{Xonics@tongji.edu.cn}
\address{
$^{1}$Department of Physics, Zhejiang Normal University, Jinhua 321004, Zhejiang , P. R. China\\
$^{2}$Center for Phononics and Thermal Energy Science, China-EU Joint Center for Nanophononics, \\
Shanghai Key Laboratory of Special Artificial Microstructure Materials and Technology,  \\
School of Physics Sciences and Engineering, Tongji University, Shanghai 200092, China
}

\date{\today}

\begin{abstract}
We investigate quantum heat transfer and thermal management in the nonequilibrium qubit-phonon hybrid system by applying the quantum master equation embedded with phononic coherent state.
We obtain the steady state heat flow by tuning the arbitrary qubit-phonon coupling strength,
which particularly exhibits the power-law scaling behavior and the turnover behavior in the weak and strong coupling regimes, respectively.
Moreover, we analyze the negative differential thermal conductance and thermal rectification, which becomes profound with weak qubit-phonon interaction
and large temperature bias.
These results would contribute to smart energy control and design of  phononic hybrid quantum devices.
\end{abstract}

%\pacs{}
%keywords
%open quantum systems; quantum heat transfer
%hybrid quantum system;
%quantum master equation; phononic coherent state
\maketitle

\section{Introduction}

%part 1
The tremendous progress of quantum engineering spurs on the generation of hybrid quantum systems(HQSs),
which establish multitasking platforms for the practical realization in versatile areas, ranging from
quantum optics, quantum information science to atomic physics~\cite{mwallquist2009pst,gkurizki2015pnas}.
HQSs are typically composed by two or more quantum components,  with each owning distinct physical functionality,
%e.g., spin-photon interface~\cite{awallraff2004nature,jqyou2011nature,akockum2019nature,pdiaz2019rmp}, optomechanic interaction~\cite{fmarquardt2009phys,maspelmeyer2014rmp} and spin-phonon coupling~\cite{aconnell2010nature}.
e.g., spin storing long-lived memory and photon transmitting high-quality information via the spin-photon interface~\cite{ablais2004pra,awallraff2004nature,tniemczyk2010np,aconnell2010nature,jqyou2011nature,lzhu2013prl}.
The main advantage of such quantum hybridization is that HQSs overcome individual limitations, and probably make the universal applications.

%part 2 CQED
The representative spin-photon hybrid system is the circuit quantum electrodynamics(cQED) platform,
which is generally composed by a superconducting qubit coupled to the on-chip microwave resonator~\cite{ablais2004pra,awallraff2004nature}.
The cQED is theoretically modeled as the Jaynes-Cummings model with weak hybrid coupling~\cite{ablais2004pra} and quantum Rabi model in the strong coupling regime~\cite{tniemczyk2010np}, respectively.
It has been extensively applied to investigate quantum correlation enhancement out-of-equilibrium~\cite{nfjohnson2015njp},
and coherent control of quantum photon transport~\cite{lzhu2013prl}
and quantum network communication~\cite{jicirac1997prl,hkimble2008nature}.
%as the building unit in the quantum network~\cite{} by combining the spin-photon coupling induced nonlinearity and the lattice induced delocalization of photons,
%resulting the significant advance of   and quantum internet.
%, the cQED-based quantum processer was built by John Martinis \emph{etc.} in Google, who claimed the realization of quantum supremacy~\cite{farute2019nature}.
%part 3 phonon, main , information and energy transfer, lack

As an analogy of photons, quantum information processing~\cite{prabl2010np,skolkowitz2012science,deassis2014nn,mjaschuetz2015prx,abienfait2019science}
and quantum logical operation~\cite{jhj2015prb,cthann2019prl} have also be widely conducted based on the qubit-phonon hybridization,
which typically consists of one two-level qubit mechanically interacting with a phonon mode, which can be realized by
single electronic qubit coupled to the nanomechanical resonator~\cite{prabl2010np} or the acoustic resonator~\cite{cthann2019prl},
one quantum dot embedded within a nanowire~\cite{deassis2014nn},
and one molecule junction coupled to the inter(intra)-molecular vibration~\cite{dsegal2016arpc}.
If the two-level qubit is replaced by the qubits ensemble, the dynamically cooperative effects can be  observed,
ranging from the fast phonon dynamics~\cite{ceban2017pra}, superradiant lasing~\cite{droenner2017pra} to ground state cooling~\cite{montenegro2018pra}.

Quantum energy flow, which is tightly related with quantum information science,
is considered as the key characteristic to detect the nonequilibrium behavior of open quantum systems~\cite{lawu2011pra,kmicadei2019nc,zxman2019qip,cllatune2019prr}.
In particular, the heat transfer has been extensively studied within the nonequilibrium qubit systems~\cite{dsegal2005prl,dsegal2008prl,jren2010prl}
and anharmonic phononic lattices~\cite{dhhe2016prb,zqzhang2017prb,dhhe2018prb}.
While for the phononic HQSs, the vibration mode mostly plays the assistant role to enhance the electron transfer\cite{jhj2015prb,dsegal2016arpc,jxzhu2003prb,jren2012prb,arrachea2014prb,agarwalla2015prb,hartle2018prb}.
Hence, considering the successful applications of qubit-phonon hybridization in the quantum information science and nonequilibrium effects,
we are motivated to exploring the quantum heat transfer in the qubit-phonon hybrid system on an equal footing.
%Moreover, in the thermodynamic limit, it can also be applied to investigate quantum thermal transport~\cite{zqzhang2017prb} based on the Holstein-Primakoff approximation.
%Hence, the qubit-phonon hybrid system provides a solid platform to investigate the quantum information and energy transport.
%Recently, the quantum information processing is preliminarily unraveled to be tightly tied to the quantum heat transfer,
%where the direction of heat flow can be reversed by the quantum correlation.
Moreover, the phononic logical operation requires smart phononic devices~\cite{nbli2012rmp},
which efficiently manage thermal energy.
Negative differential thermal conductance(NDTC), phononic rectifier and transistor are considered as key functional components~\cite{jhj2015prb,bwli2004prl,bwli2006apl,dhhe2009prb,dhhe2010pre,hkchan2014pre}.
However, the investigation of quantum dot-phonon hybrid rectifier and transistor is limited to the linear response regime,
and the heat flow is correlated with the electric current~\cite{jhj2015prb}.
While the phonon-lattice transistor mainly works in the classical regime~\cite{nbli2012rmp}.
%and the heat current is accompanied by the electric current.
Hence, it is demanding to investigate thermal management far-from equilibrium in the phononic HQSs.

% part 4 we study quantum heat transport NDTC, recticfication, amplification
In this paper, to give a theoretic view of the  quantum energy transfer in the nonequilibrium qubit-phonon hybrid system,
we mainly study the steady state heat flow and  thermal management, which is driven by temperature bias.
We apply the quantum master equation combined with the phononic coherent state to derive the dynamical equation of the hybrid system density matrix in Sec. II.
The inclusion of the coherent state enables us to properly treat non-weak qubit-phonon coupling.
In Sec. III, we obtain the analytical expression of the steady state heat current, and the scaling behavior in the weak qubit-phonon coupling limit
is analytically estimated.
In Sec. IV, we investigate the representative effects of thermal management, i.e. NDTC~\cite{bwli2006apl,dhhe2009prb,dhhe2010pre,hkchan2014pre} and thermal rectification~\cite{bwli2004prl}.
%and heat amplification~\cite{nbli2012rmp}.
The underlying mechanism of the NDTC and the connection between these effects are discussed.
Finally, we give a brief summary in Sec. V.

\section{Model and method}
\subsection{Nonequilibrium qubit-phonon hybrid model}
The nonequilibrium qubit-phonon hybrid system consisting of one qubit coupled to a single mode phononic field, each interacting with an individual thermal bath,
$\hat{H}=\hat{H}_s+\sum_{u=q,c}(\hat{H}^u_{b}+\hat{V}_u)$.
Specifically, the system Hamiltonian is described as~\cite{prabl2010np}
\begin{eqnarray}~\label{hs}
\hat{H}_s=\frac{\varepsilon}{2}\hat{\sigma}_z+\omega_0\hat{a}^{\dag}\hat{a}+\lambda\hat{\sigma}_z(\hat{a}^{\dag}+\hat{a}),
\end{eqnarray}
where
$\hat{\sigma}_z=|\uparrow{\rangle}{\langle}\uparrow|-|\downarrow{\rangle}{\langle}\downarrow|$ are Pauli operators
with the excited(ground) state $|\uparrow{\rangle}~(|\downarrow{\rangle})$ of the qubit,
$\varepsilon$ is the Zeeman splitting energy,
$\hat{a}^{\dag}~(\hat{a})$ creates(annihilates) one phonon with the frequency $\omega_0$,
and $\lambda$ is the interaction strength between the qubit and the bosonic field.
The $u$th thermal bath is described as
$\hat{H}^u_b=\sum_k\omega_k\hat{b}^{\dag}_{k,u}\hat{b}_{k,u}$,
with $\hat{b}^{\dag}_{k,u}~(\hat{b}_{k,u})$ creating(annihilating) one phonon with the frequency $\omega_k$.
The qubit-bath interaction is given by
\begin{eqnarray}~\label{vq}
\hat{V}_q=\hat{\sigma}_x\sum_k(f_{k,q}\hat{b}^{\dag}_{k,q}+f^{*}_{k,q}\hat{b}_{k,q}),
\end{eqnarray}
and the phonon-bath interaction is given by
\begin{eqnarray}~\label{vc}
\hat{V}_c=\sum_k(f_{k,c}\hat{b}^{\dag}_{k,q}\hat{a}+f^{*}_{k,c}\hat{a}^{\dag}\hat{b}_{k,q}),
\end{eqnarray}
with $\hat{\sigma}_x=|\uparrow{\rangle}{\langle}\downarrow|+|\downarrow{\rangle}{\langle}\uparrow|$   and $f_{k,q(c)}$ the corresponding coupling strength.
The $u$th thermal bath is characterized as the spectral function
$\gamma_u(\omega)=2\pi\sum_k|f_{k,u}|^2\delta(\omega-\omega_k)$, which is specified as the Ohmic form
$\gamma_u(\omega)=\alpha_u\omega\exp(-|\omega|/\omega_{c,u})$,
with the coupling strength $\alpha_u$ and the cutoff frequency $\omega_{c,u}$~\cite{uweiss2012book}.
In the following, we select $\omega_0$ as the energy unit for convenience without losing any generality.

For the qubit-phonon interacting system at Eq.~(\ref{hs}), it can be exactly solved by applying the coherent bosonic state due to the commutating relation $[\hat{\sigma}_z,\hat{H}_s]=0$.
Specifically, by projecting $\hat{H}_s$ to the qubit states $|\uparrow{\rangle}$ and $|\downarrow{\rangle}$,
we obtain
$\hat{H}_s|\uparrow{\rangle}=
[\omega_0(\hat{a}^{\dag}+\lambda/\omega_0)(\hat{a}+\lambda/\omega_0)-\lambda^2/\omega_0+\varepsilon/2]|\uparrow{\rangle}$
and
$\hat{H}_s|\downarrow{\rangle}=
[\omega_0(\hat{a}^{\dag}-\lambda/\omega_0)(\hat{a}-\lambda/\omega_0)-\lambda^2/\omega_0-\varepsilon/2]|\downarrow{\rangle}$.
Hence, the eigenstates are given by
$|\phi^{\uparrow}_n{\rangle}=|n{\rangle}_{\uparrow}{\otimes}|\uparrow{\rangle}$
and $|\phi^{\downarrow}_n{\rangle}=|n{\rangle}_{\downarrow}{\otimes}|\downarrow{\rangle}$,
where the phononic coherent states are
\begin{subequations}
\begin{align}
|n{\rangle}_{\uparrow}=&\frac{(\hat{a}^{\dag}+\lambda/\omega)^n}{\sqrt{n!}}\exp[-(\lambda/\omega_0)^2-(\lambda/\omega_0)\hat{a}^{\dag}]|0{\rangle}_a,~\label{state1}\\
|n{\rangle}_{\downarrow}=&\frac{(\hat{a}^{\dag}-\lambda/\omega)^n}{\sqrt{n!}}\exp[-(\lambda/\omega_0)^2+(\lambda/\omega_0)\hat{a}^{\dag}]|0{\rangle}_a,~\label{state11}
\end{align}
\end{subequations}
with the Fock state in vacuum $\hat{a}|0{\rangle}_a=0$.
And the eigenvalues are given by
\begin{subequations}
\begin{align}
E_{n,\uparrow}=&\omega_0n-\lambda^2/\omega_0+\varepsilon/2\\
E_{n,\downarrow}=&\omega_0n-\lambda^2/\omega_0-\varepsilon/2.
\end{align}
\end{subequations}
\begin{comment}
For light-matter interacting systems, the seminal model is the quantum Rabi model,
where the interaction between the qubit and the photon is expressed as
$\hat{V}_{\textrm{Rabi}}=\lambda\hat{\sigma}_x(\hat{a}^{\dag}+\hat{a})$,
in which the energy transfer is accompanied by the spin flip and photon emission(absorption).
It has been realized in the circuit QED, where a spin $1/2$ is transversely coupled to a single bosonic mode,
referred as XcQED~\cite{awallraff2004nature}.
While for the qubit-boson interaction in Eq.~(\ref{hs}), it can be realized via the longitudinal coupling~\cite{pmb2015prb,sricher2017prb}.
Moreover, such interaction can be unravelled in quantum dots coupled to a nanomechanical oscillator~\cite{vceban2017pra}.
\end{comment}

%The qubit-phonon hybrid system at Eq.~(\ref{hs}) can be realized in single trapped-ion embedded in long-wavelength radiation~\cite{fmintert2001prl},
%one quantum dot coupled to a membrane~\cite{vceban2017pra} or high-Q acoustic cavity~\cite{ldroenner2017pra},
%and one spin interacting with a nanomechanical oscillator~\cite{vmontenegro2018pra}.
%Moreover, as the high frequency analogy, such qubit-phonon interaction can also be simulated by
%the Xc-QED, where a $1/2$-spin is  longitudinal coupled to a single photonic mode~\cite{pmb2015prb,sricher2017prb}.

%%==========================================
\begin{figure}[tbp]
\begin{center}
%\vspace{-1.0cm}
\includegraphics[scale=0.3]{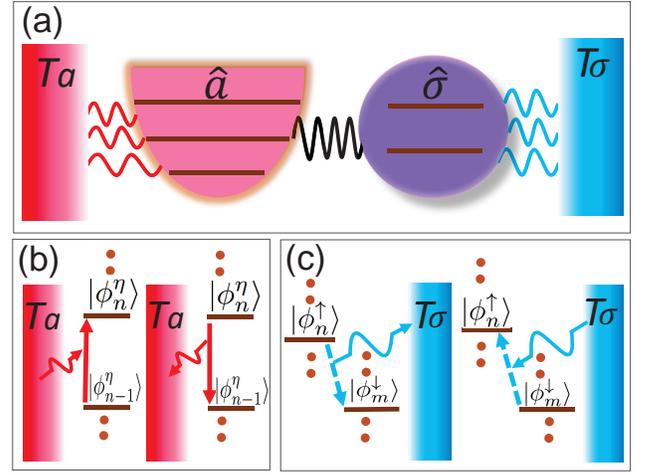}
%\vspace{-2.0cm}
\end{center}
\caption{(Color online)
(a) Schematic illustration of single mode phononic field(pink half-circle marked with $\hat{a}$)
interacting with a two-level qubit(blue circle marked with $\hat{\sigma}$),
each individually coupled to a thermal bath characterized as the temperature $T_a$ and $T_{\sigma}$,
and the dark wave line describes the interaction between the  phononic field and the qubit;
(b) transitions between the eigenstates $|\phi^{\eta}_{n}{\rangle}$ and $|\phi^{\eta}_{n-1}{\rangle}$ assisted by the $a$th bath described by the rates at Eq.~(\ref{gap}) and Eq.~(\ref{gam}),
with $\eta=\uparrow,\downarrow$ and $n{\geq}1$;
(c) transitions between the eigenstates $|\phi^{\uparrow}_{n}{\rangle}$ and $|\phi^{\downarrow}_{m}{\rangle}$ assisted by the $\sigma$th bath described by the rates at Eq.~(\ref{gamma1}) and Eq.~(\ref{gamma2}), and the coefficient at Eq.~(\ref{s1}).
}~\label{fig1}
\end{figure}
%%==========================================

\subsection{Quantum master equation}
We apply the quantum master equation to study dissipative dynamics of the qubit-phonon interacting system.
We consider weak system-bath interactions in Eq.~(\ref{vq}) and Eq.~(\ref{vc}), where the Born-Markov approximation becomes applicable.
Accordingly, the total density matrix is decomposed as
$\hat{\rho}_{\textrm{tot}}(t){\approx}\hat{\rho}_s(t){\otimes}\hat{\rho}_b$, where $\hat{\rho}_s(t)$ is the reduced density operator of the hybrid system
and $\hat{\rho}_b=\exp(-\sum_{u=q,c}\hat{H}^u_b/k_BT_u)/\textrm{Tr}\{\exp(-\sum_{u=q,c}\hat{H}^u_b/k_BT_u)\}$ the equilibrium distribution of thermal baths,
with $k_B$ the Boltzmann constant and $T_u$ the temperature of the $u$th thermal bath.
Then, by perturbing Eq.~(\ref{vq}) and Eq.~(\ref{vc}) in the eigenspace of $\hat{H}_s$ up to the second order separately, we obtain the
nonequilibrium dressed master equation~\cite{fbeaudoin2011pra,alboite2016pra}
\begin{eqnarray}~\label{dme1}
\frac{d\hat{\rho}_s}{dt}&=&-i[\hat{H}_s,\hat{\rho}_s]
+\sum_{n,m;\eta}
\{\Gamma^+_q(\phi^{\overline{\eta}}_m|\phi^\eta_n)\mathcal{D}_q[|\phi^\eta_n{\rangle}{\langle}\phi^{\overline{\eta}}_m|]\hat{\rho}_s\nonumber\\
&&+\Gamma^-_q(\phi^\eta_n|\phi^{\overline{\eta}}_m)\mathcal{D}_{q}[|\phi^{\overline{\eta}}_m{\rangle}{\langle}\phi^\eta_n|]\hat{\rho}_s\}\nonumber\\
&&+\sum_{n,m;\eta}
\{\Gamma^+_c(\phi^\eta_m|\phi^\eta_n)\mathcal{D}_c[|\phi^\eta_n{\rangle}{\langle}\phi^\eta_m|]\hat{\rho}_s\nonumber\\
&&+\Gamma^-_c(\phi^\eta_n|\phi^\eta_m)\mathcal{D}_{c}[|\phi^\eta_m{\rangle}{\langle}\phi^\eta_n|]\hat{\rho}_s\}
\end{eqnarray}
where $\overline{\eta}=\uparrow(\downarrow)$ for $\eta=\downarrow(\uparrow)$,
the dissipator involved with the $u$th thermal bath is given by
\begin{eqnarray}
\mathcal{D}_u[|\phi^{\eta^{\prime}}_m{\rangle}{\langle}\phi^\eta_n|]\hat{\rho}_s&=&
|\phi^{\eta^{\prime}}_m{\rangle}{\langle}\phi^\eta_n|\hat{\rho}_s|\phi^\eta_n{\rangle}{\langle}\phi^{\eta^{\prime}}_m|\\
&&-\frac{1}{2}(|\phi^\eta_n{\rangle}{\langle}\phi^\eta_n|\hat{\rho}_s
+\hat{\rho}_s|\phi^\eta_n{\rangle}{\langle}\phi^\eta_n|),\nonumber
\end{eqnarray}
and the transition rates are
\begin{subequations}
\begin{align}
\Gamma^{+}_u(\phi^{\eta^{\prime}}_m|\phi^\eta_n)=&\theta(\Delta^{m,\eta^{\prime}}_{n,\eta})
\gamma_u(\Delta^{m,\eta^\prime}_{n,\eta})n_u(\Delta^{m,\eta^\prime}_{n,\eta})
|{\langle}\phi^\eta_n|\hat{S}^{\dag}_u|\phi^{\eta^{\prime}}_m{\rangle}|^2,~\label{gamma1}\\
\Gamma^{-}_u(\phi^{\eta^{\prime}}_m|\phi^\eta_n)=&\theta(\Delta^{n,\eta}_{m,\eta^\prime})
\gamma_u(\Delta^{n,\eta}_{m,\eta^\prime})[1+n_u(\Delta^{n,\eta}_{m,\eta^\prime})]~\label{gamma2}\nonumber\\
&{\times}|{\langle}\phi^{\eta^{\prime}}_m|\hat{S}_u|\phi^\eta_n{\rangle}|^2,
\end{align}
\end{subequations}
with the herald function $\theta(x)=1$ for $x>0$ and $\theta(x)=0$ for $x{\le}0$,
the energy gap $\Delta^{m,\eta^{\prime}}_{n,\eta}=E_{n,\eta}-E_{m,\eta^{\prime}}$,
and the Bose-Einstein distribution function
$n_u(\omega)=1/[\exp(\omega/k_BT_u)-1]$.
The system operators are $\hat{S}_q=\hat{\sigma}_x$
and $\hat{S}_c=\hat{a}$, where the nonzero transition coefficients are given by
\begin{subequations}
\begin{align}
{\langle}\phi^\eta_n|\hat{a}^\dag|\phi^\eta_m{\rangle}=&\sqrt{m+1}\delta_{n,m+1}-g_\eta\delta_{n,m},~\label{a1}\\
{\langle}\phi^{\uparrow}_n|\hat{\sigma}_x|\phi^{\downarrow}_m{\rangle}=&(-1)^nD_{nm}(2\lambda/\omega),~\label{s1}
\end{align}
\end{subequations}
with ${\langle}\phi^{\downarrow}_n|\hat{\sigma}_x|\phi^{\uparrow}_m{\rangle}=(-1)^mD_{nm}(2\lambda/\omega)$, $g_{\uparrow}=\lambda/\omega_0$, $g_{\downarrow}=-\lambda/\omega_0$
and the coefficient~\cite{qhchen2008pra}
\begin{eqnarray}~\label{dnm-def}
D_{nm}(x)=e^{-x^2/2}\sum^{\min[n,m]}_{l=0}\frac{(-1)^l\sqrt{n!m!}x^{n+m-2l}}{(n-l)!(m-l)!l!}.
\end{eqnarray}
The rate $\Gamma^{+(-)}_u(\phi^{\eta^\prime}_m|\phi^\eta_n)$ in Eq.~(\ref{gamma1})[Eq.~(\ref{gamma2})] describes the transition
from the eigenstate $|\phi^{\eta^\prime}_m{\rangle}$ up(down) to $|\phi^\eta_n{\rangle}$,
which is assisted by absorbing(emitting) one phonon with energy $\Delta^{m,\eta^{\prime}}_{n,\eta}>0$ from(into) the $u$th bath.

It is interesting to find that the nonzero transition rates assisted by the $a$th bath at Eq.~(\ref{gamma1}) and Eq.~(\ref{gamma2})
\begin{subequations}
\begin{align}
\Gamma^{+}_a(\phi^\eta_{m}|\phi^\eta_{m+1})=&\gamma_a(\omega_0)n_a(\omega_0)(m+1),~\label{gap}\\
\Gamma^{-}_a(\phi^\eta_{m+1}|\phi^\eta_{m})=&\gamma_a(\omega_0)[1+n_a(\omega_0)](m+1),~\label{gam}
\end{align}
\end{subequations}
are irrelevant with the qubit-phonon coupling strength, shown in Fig.~\ref{fig1}(b).
They obey the detailed balance relationship
$\Gamma^{+}_a(\phi^\eta_{m}|\phi^\eta_{m+1})/\Gamma^{-}_a(\phi^\eta_{m}|\phi^\eta_{m+1})=\exp(-\omega_0/k_BT_a)$.
While $\Gamma^{\pm}_\sigma(\phi^{\eta^{\prime}}_{m}|\phi^\eta_{n})$ describes the transition between eigenstates
$|\phi^{\eta^{\prime}}_{m}{\rangle}$ and $|\phi^\eta_{n}{\rangle}$ assisted by the $\sigma$th bath, shown in Fig.~\ref{fig1}(c).
It strongly depends on the coupling strength
reflected by the coefficient $D_{nm}(2\lambda/\omega_0)$.
Moreover, it should be noted that the  quantum master equation we applied in this paper includes
the transition between eigenstates, i.e. the dressed state $|\phi^{\uparrow(\downarrow)}_n{\rangle}$.
Hence, we are able to investigate system dynamics with strong coupling between the qubit and the phononic field,
which can be comparable with the characteristic energy unit $\omega_0$.
%~\cite{fbeaudoin2011pra}.
%While for the traditional optical master equation, it fails to capture the dressed picture of the dissipative system,
%i.e. $\hat{S}(\tau){\approx}e^{i\hat{H}^0_s\tau}\hat{S}e^{-i\hat{H}^0_s\tau}$ with
%$\hat{S}=\hat{a},\hat{\sigma}_x$ and
%$\hat{H}^0_s=\frac{\varepsilon}{2}\hat{\sigma}_z+\omega_0\hat{a}^\dag\hat{a}$.
%Therefore, it is valid only in  the weak qubit-phonon coupling  limit.

%%==========================================
\begin{figure}[tbp]
\begin{center}
%\vspace{-1.0cm}
\includegraphics[scale=0.4]{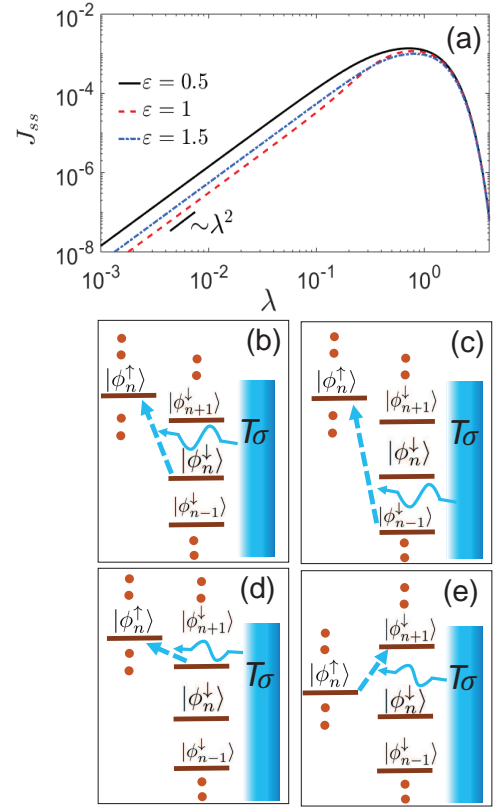}
%\vspace{-2.0cm}
\end{center}
\caption{(Color online)
(a) Steady state heat current $J_{ss}$ as a function of the qubit-phonon coupling strength $\lambda$;
microscopic transitions in the weak qubit-phonon coupling limit:(b),(c) described by the rate $\Gamma^+_{\sigma}(\phi^{\downarrow}_m|\phi^\uparrow_n)$ at Eq.~(\ref{gpm-weak}),
and (d), (e) described by the rate $\Gamma^+_{\sigma}(\phi^{\downarrow}_{n+1}|\phi^\uparrow_n)$ and $\Gamma^+_{\sigma}(\phi^{\uparrow}_{n}|\phi^\downarrow_{n+1})$ at Eq.~(\ref{gpm-weak})
in the regimes $\varepsilon{>}\omega_0$ and  $\varepsilon{<}\omega_0$, respectively.
The other parameters are given by $\omega_0=1$, $\alpha_a=\alpha_\sigma=0.005$, and $\omega_c=10$, $T_a=1.5$, and $T_\sigma=0.5$.
}~\label{fig2}
\end{figure}
%%==========================================
\section{Steady state heat current}

%{General expression}
Under the temperature bias$(T_a{\neq}T_\sigma)$, a heat flow naturally occurs mediated by the qubit-phonon system at steady state.
We apply the  dressed master equation at Eq.~(\ref{dme1}) to obtain the steady state heat current.
Specifically, the density matrix of the hybrid system $P_{n,\eta}={\langle}\phi^\eta_n|\hat{\rho}_s|\phi^\eta_n{\rangle}$ is given by
\begin{eqnarray}~\label{dpn}
\frac{dP_{n,\eta}}{dt}&=&\sum_{u;m,\eta^\prime}
[\Gamma^{+}_u(\phi^{\eta^\prime}_m|\phi^\eta_n)P_{m,\eta^\prime}
-\Gamma^{+}_u(\phi^\eta_n|\phi^{\eta^\prime}_m)P_{n,\eta}]\\
&&+\sum_{u;m,\eta^\prime}
[\Gamma^{-}_u(\phi^{\eta^\prime}_m|\phi^\eta_n)P_{m,\eta^\prime}
-\Gamma^{-}_u(\phi^\eta_n|\phi^{\eta^\prime}_m)P_{n,\eta}].\nonumber
\end{eqnarray}
By analyzing the state transitions at Eq.~(\ref{dpn}), the steady state heat current into the $\mu$th thermal bath can be obtained as
\begin{eqnarray}~\label{current1}
J_\mu&=&\sum_{n,m;\eta,\eta^\prime}\Delta^{m,\eta^\prime}_{n,\eta}
\Gamma^-_\mu(\phi^\eta_n|\phi^{{\eta}^\prime}_m)P^{ss}_{n,{\eta}}\\
&&-
\sum_{n,m;\eta,\eta^\prime}
\Delta^{n,\eta}_{m,\eta^\prime}\Gamma^+_\mu(\phi^\eta_n|\phi^{{\eta}^\prime}_m)P^{ss}_{n,{\eta}}.\nonumber
\end{eqnarray}
where the first(second) term describes the heat transfer into(out of) the $\mu$th bath via the energy down(up) transition from
the state $|\phi^{\eta}_n{\rangle}$ to $|\phi^{\eta^\prime}_m{\rangle}$
by releasing(absorbing) energy $\Delta^{m,\eta^\prime}_{n,\eta}$.
In the following, we define the steady state current as
\begin{eqnarray}
J_{ss}=J_{\sigma}=-J_{a},
\end{eqnarray}
where the energy conservation $J_{\sigma}+J_{a}=0$ is fulfilled, which can be verified from Eq.~(\ref{dpn}) and Eq.~(\ref{current1}).

Then, we investigate the effect of the qubit-phonon coupling strength $\lambda$ on the steady state heat current $J_{ss}$ by numerically plotting Fig.~\ref{fig2}(a).
It is found that the steady state current shows the optimal behavior by tuning  $\lambda$ over a wide coupling regime.
Specifically, in the weak qubit-phonon interaction regime $J_{ss}$ is enhanced by increasing the coupling strength $\lambda$,
%Moreover, heat current shows the scaling behavior $J_{ss}{\propto}\lambda^2/\omega^2_0$,
%which will be analytically estimated in the following subsection.
While in the strong qubit-phonon coupling regime, the heat current is monotonically suppressed.
%It mainly comes from the fact that the dressed-state transitions between $|\phi^{\uparrow}_n{\rangle}$ and $|\phi^{\downarrow}_m{\rangle}$ assisted by the spin flip are strongly blocked with large coefficient $\lambda/\omega_0$[i.e. $D_{nm}(2\lambda/\omega_0){\ll}1$ with the dominant factor $\exp{(-\lambda^2/2\omega^2_0)}{\ll}1$].
It should be admitted that it is difficult to  obtain an explicit expression of the steady state heat current with arbitrary coupling strength.
In the following, we try to analytically explore the steady state behavior of the heat current in the weak and strong coupling limits, respectively.

\subsection{Weak qubit-phonon coupling limit}
%Here, we focus on the heat flow with the weak qubit-phonon interaction.

As $\lambda/\omega_0{\ll}1$, the coherent state coefficient is approximately reduced to
\begin{eqnarray}~\label{dnmweak}
D_{nm}(\frac{2\lambda}{\omega_0})&{\approx}&(-1)^n\delta_{n,m}
+\frac{2\lambda}{\omega_0}(-1)^n\sqrt{n+1}\delta_{n,m-1}\nonumber\\
&&+\frac{2\lambda}{\omega_0}(-1)^{n-1}\sqrt{n}\delta_{n,m+1}.
\end{eqnarray}
Consequently, the transition rates  $\Gamma^{\pm}_\sigma(\phi^{\downarrow}_{m}|\phi^\uparrow_{n})$  are reduced to
%\begin{eqnarray}~\label{gpm-weak}
%\Gamma^{\pm}_\sigma(\phi^{\downarrow}_{m}|\phi^\uparrow_{n})&\approx&
%[\kappa^{\pm}_\sigma(\varepsilon)
%-(\frac{2\lambda}{\omega_0})\delta_{n,m+1}\sqrt{n}\kappa^{\pm}_\sigma(\varepsilon+\omega_0)\\
%&&+(\frac{2\lambda}{\omega_0})\delta_{n,m-1}\sqrt{m}\theta(\varepsilon-\omega_0)\kappa^{\pm}_\sigma(\varepsilon-\omega_0)],\nonumber
%\end{eqnarray}
\begin{eqnarray}~\label{gpm-weak}
\Gamma^{\pm}_\sigma(\phi^{\overline{\eta}}_{m}|\phi^\eta_{n})&\approx&
\delta_{\eta,\uparrow}\kappa^{\pm}_\sigma(\varepsilon)+(\frac{2\lambda}{\omega_0})^2\delta_{n,m-1}{m}{\times}\nonumber\\
&&[\delta_{\eta,\uparrow}\kappa^{\pm}_\sigma(\varepsilon-\omega_0)+\delta_{\eta,\downarrow}\kappa^{\pm}_\sigma(\omega_0-\varepsilon)]\nonumber\\
&&-(\frac{2\lambda}{\omega_0})^2\delta_{n,m+1}\delta_{\eta,\uparrow}{n}\kappa^{\pm}_\sigma(\varepsilon+\omega_0),
\end{eqnarray}
where the sequential rates are $\kappa^+_{\sigma}(\omega)=\theta(\omega)\gamma_\sigma(\omega)n_\sigma(\omega)$
and
$\kappa^-_{\sigma}(\omega)=\theta(\omega)\gamma_\sigma(\omega)[1+n_\sigma(\omega)]$,
with $\theta(x>0)=1$ and $\theta(x{\leq}0)=0$.
Hence, it is known that the energy transport associated with the qubit flip is determined by two kinds of transition processes:
1)eigenstate transition in absence of phonon hopping $|\phi^{\uparrow}_m{\rangle}{\leftrightarrows}|\phi^{\downarrow}_m{\rangle}$ in Fig.~\ref{fig2}(b),
2)eigenstate  transitions involved with phonon transfer processes $|\phi^{\uparrow}_{m+1}{\rangle}{\leftrightarrows}|\phi^{\downarrow}_m{\rangle}$
and $|\phi^{\uparrow}_{m}{\rangle}{\leftrightarrows}|\phi^{\downarrow}_{m+1}{\rangle}$ in Figs.~\ref{fig2}(c-e).
In particular, the second type process is crucial to exhibit the steady state heat flow by both including the qubit flip and phonon hopping simultaneously.

Then, if we reorganize the populations in the vector form
$|\mathbf{P}{\rangle\rangle}=[P^{ss}_{0,\uparrow},P^{ss}_{1,\uparrow},\cdots,P^{ss}_{0,\downarrow},P^{ss}_{1,\downarrow},\cdots]^T$
with the index $T$ the transpose of the vector,
the dynamical equation is re-expressed as[see Appendix A for the detail information of the dynamical equations and transition matrix elements]
\begin{eqnarray}
\frac{d}{dt}|\mathbf{P}{\rangle\rangle}{\approx}(\mathbf{M}_{a}+\mathbf{M}_{\sigma})|\mathbf{P}{\rangle\rangle}
+(2\lambda/\omega_0)^2\mathbf{M}_{\lambda}|\mathbf{P}{\rangle\rangle}.
\end{eqnarray}
$\mathbf{M}_{a}$ describes transitions between $|\phi^\eta_{m}{\rangle}$ and $|\phi^\eta_{m{\pm}1}{\rangle}$ assisted by the phonon excitation(annihilation), characterized by the transition rates
$\Gamma^+_a(\phi^\eta_{m}|\phi^\eta_{m+1})$ and $\Gamma^-_a(\phi^\eta_{m}|\phi^\eta_{m-1})$;
$\mathbf{M}_{\sigma}$  describes transitions between $|\phi^\eta_{m}{\rangle}$ and $|\phi^{\overline{\eta}}_{m}{\rangle}$, which is assisted
by the spin flip and characterized by the rates
$\Gamma^+_\sigma(\phi^\downarrow_{m}|\phi^\uparrow_{m})$ and $\Gamma^-_\sigma(\phi^\uparrow_{m}|\phi^\downarrow_{m})$;
$\mathbf{M}_{\lambda}$ describes transitions $|\phi^\uparrow_{m}{\rangle}{\leftrightarrow}|\phi^\downarrow_{m-1}{\rangle}$
and $|\phi^\uparrow_{m-1}{\rangle}{\leftrightarrow}|\phi^\downarrow_{m}{\rangle}$ cooperatively contributed by the qubit flip and phonon hopping,
which are characterized by rates $\Gamma^+_\sigma(\phi^\downarrow_{m}|\phi^\uparrow_{m{\pm}1})$
and $\Gamma^-_\sigma(\phi^\uparrow_{m}|\phi^\downarrow_{m{\pm}1})$.
At steady state $\frac{d}{dt}|\mathbf{P}{\rangle\rangle}_{ss}=0$, the state state solution is obtained as
\begin{eqnarray}
|\mathbf{P}{\rangle\rangle}_{ss}{\approx}|\mathbf{P}_{(0)}{\rangle\rangle}-(\frac{2\lambda}{\omega_0})^2
\hat{Q}(\mathbf{M}_{a}+\mathbf{M}_{\sigma})^{-1}\hat{Q}\mathbf{M}_{\lambda}|\mathbf{P}_{(0)}{\rangle\rangle},
\end{eqnarray}
where the projecting operator is $\hat{Q}=1-|\mathbf{P}_{(0)}{\rangle\rangle}{\langle\langle}\textbf{I}|$, the unit vector is
${\langle\langle}\textbf{I}|\mathbf{P}_{(0)}{\rangle\rangle}=1$,
and $|\mathbf{P}_{(0)}{\rangle\rangle}$ is the solution $(\mathbf{M}_{a}+\mathbf{M}_{\sigma})|\mathbf{P}_{(0)}{\rangle\rangle}=0$,
with the corresponding steady state distribution of the hybrid system
\begin{eqnarray}~\label{rho00}
\hat{\rho}^{(0)}_{ss}=\frac{\exp[-{\varepsilon}\hat{\sigma}_z/(2k_BT_q)-\omega_0\hat{a}^{\dag}\hat{a}/(k_BT_c)]}
{(2\cosh[\varepsilon/(2k_BT_q)][1+n_a(\omega_0)])}.
\end{eqnarray}

Moreover, the current at Eq.~(\ref{current1}) is expressed as
\begin{eqnarray}~\label{ja1}
J_{ss}=\omega_0\sum_{m,\eta}m[\kappa^+_a(\omega_0)P^{ss}_{m-1,\eta}-\kappa^-_{a}(\omega_0)P^{ss}_{m,\eta}].
\end{eqnarray}
%is approximated as
%\begin{widetext}
%\begin{eqnarray}
%J_{ss}&\approx&\varepsilon\sum_m[\kappa^-_\sigma(\varepsilon)P^{ss}_{m,\uparrow}-\kappa^+_\sigma(\varepsilon)P^{ss}_{m,\downarrow}]\\
%&&+(\omega_0+\varepsilon)(2\lambda/\omega_0)^2\sum_{m}
%m[\kappa^-_\sigma(\omega_0+\varepsilon)P^{ss}_{m,\uparrow}-\kappa^+_\sigma(\varepsilon+\omega_0)P^{ss}_{m-1,\downarrow}]\nonumber\\
%&&+(\omega_0-\varepsilon)(2\lambda/\omega_0)^2\sum_{m}
%m[\kappa^+_\sigma(\omega_0-\varepsilon)P^{ss}_{m-1,\uparrow}-\kappa^-_\sigma(\omega_0-\varepsilon)P^{ss}_{m,\downarrow}].\nonumber
%\end{eqnarray}
%\end{widetext}
Considering the steady state condition in absence of the qubit-phonon interaction($\lambda=0$)
$\sum_m[\kappa^-_\sigma(\varepsilon)P^{(0)}_{m,\uparrow}-\kappa^+_\sigma(\varepsilon)P^{(0)}_{m,\downarrow}]=0$,
with $P^{(0)}_{m,\eta}$
the element of $\hat{\rho}^{(0)}_{ss}$ at Eq.~(\ref{rho00}),
it is clearly shown that in the weak qubit-phonon interaction limit, the steady state heat current scales as
\begin{eqnarray}
J_{ss}{\propto}(\lambda/\omega_0)^2.
\end{eqnarray}

\subsection{Strong qubit-phonon coupling limit}
When $\lambda/\omega_0{\gg}1$, the coherent state coefficient $D_{nm}(2\lambda/\omega_0)$ at Eq.~(\ref{dnm-def}) is strongly suppressed with increase of the coupling strength,
mainly due to the factor $\exp(-\lambda^2/2\omega^2_0){\approx}0$.
It results in the ignored transition rate assisted by the $\sigma$th thermal bath[i.e. $\Gamma^{\pm}_{\sigma}(\phi^{\eta^\prime}_m|\phi^\eta_n){\approx}0$].
Hence, the steady state populations are determined by the transition rates at Eq.~(\ref{gap}) and Eq.~(\ref{gam}), obtained as
\begin{eqnarray}
P^{ss}_{n,\uparrow}&{\approx}&\frac{\exp[-(n\omega_0+\varepsilon/2)/k_BT_a]}{2\cosh[\varepsilon/(2k_BT_a)][1+n_a(\omega_0)]},\\
P^{ss}_{n,\downarrow}&{\approx}&\frac{\exp[-(n\omega_0-\varepsilon/2)/k_BT_a]}{2\cosh[\varepsilon/(2k_BT_a)][1+n_a(\omega_0)]}.
\end{eqnarray}
Finally, the heat current is approximated as
\begin{eqnarray}
J_{ss}&\approx&\sum_{n,m}\frac{D^{2}_{nm}(\frac{2\lambda}{\omega_0})\Delta^{m,\downarrow}_{n,\uparrow}}
{2\cosh[\varepsilon/(2k_BT_a)][1+n_a(\omega_0)]}{\times}\nonumber\\
&&\{\theta(\Delta^{n,\uparrow}_{m,\downarrow})\gamma_\sigma{(\Delta^{n,\uparrow}_{m,\downarrow})}
[1+2n_\sigma{(\Delta^{n,\uparrow}_{m,\downarrow})}]e^{-\frac{n\omega_0+\varepsilon/2}{k_BT_a}}\nonumber\\
&&-\theta(\Delta^{m,\downarrow}_{n,\uparrow})\gamma_\sigma(\Delta^{m,\downarrow}_{n,\uparrow})
[1+2n_\sigma(\Delta^{m,\downarrow}_{n,\uparrow})]e^{-\frac{m\omega_0-\varepsilon/2}{k_BT_a}}\}\nonumber\\
\end{eqnarray}
which is dramatically suppressed in the strong qubit-phonon coupling limit.

%\section{Results and discussions}
\section{Thermal management}
%%==========================================
\begin{figure}[tbp]
\begin{center}
%\vspace{-1.0cm}
\includegraphics[scale=0.5]{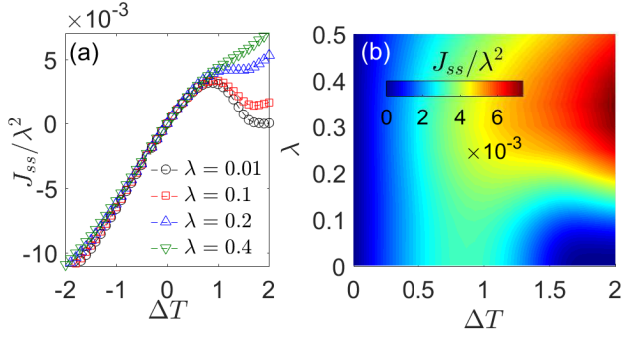}
%\vspace{-2.0cm}
\end{center}
\caption{(Color online)
(a) Steady state heat current $J_{ss}/\lambda^2$ as a function of the temperature bias ${\Delta}T=T_{a}-T_{\sigma}$
with $T_{a}=T_0+{\Delta}T/2$, $T_{\sigma}=T_0-{\Delta}T/2$ and $T_0=1$;
(b) the bird-view of the heat current $J_{ss}/\lambda^2$ by tuning both ${\Delta}T$ and $\lambda$.
The other parameters are given by $\omega_0=1$, $\varepsilon=1$, $\alpha_a=\alpha_\sigma=0.005$, and $\omega_c=10$.
}~\label{fig3}
\end{figure}
%%==========================================
\subsection{Negative differential thermal conductance}
NDTC is a typical nonlinear effect within the two baths setup, where the heat flow $J_{ss}$ is suppressed by increasing the temperature bias ${\Delta}T=T_a-T_\sigma$~\cite{dhhe2009prb,dhhe2010pre,hkchan2014pre}.
We investigate the NDTC by tuning qubit-phonon interaction strength from weak to strong in Fig.~\ref{fig3}(a).
In the positive temperature bias regime(${\Delta}T>0$),
it is interesting to find that with weak qubit-phonon coupling(e.g., $\lambda=0.01$),
the heat current shows linear increase in small ${\Delta}T$;
whereas it is monotonically suppressed with large temperature bias,
and becomes nearly vanished as ${\Delta}T=2$(corresponding to $T_a=2$ and $T_\sigma=0$).
Hence, it clearly demonstrates the emergence of the NDTC.
However, such suppression feature of the heat current is gradually weakened as the qubit-phonon interaction increases(e.g., $\lambda=0.1$),
and the nonmonotonic behavior of $J_{ss}$ disappears in the strong qubit-phonon coupling regime(e.g., $\lambda=0.2, 0.4$).
While in the negative temperature bias regime(${\Delta}T<0$), the magnitude of heat current shows monotonic enhancement by
increasing $|{\Delta}T|$, in absence of the signature of the NDTC.
To give a comprehensive picture of the NDTC, we plot a 3D view of $J_{ss}/\lambda^2$ in Fig.~\ref{fig3}(b) by both modulating ${\Delta}T$ and $\lambda$.
The existence of the NDTC is approximately limited to the coupling zone $\lambda{\in}(0,0.1)$.
Therefore, we conclude the NDTC can be exhibited under conditions of positively large temperature bias between two thermal baths and the weak qubit-phonon interaction regime.

It should be noted that though we analyze the NDTC at resonance, it can also be generally observed for the biased case(see appendix B and Fig.~\ref{fig8} for details).
Moreover, based on the NDTC, the heat amplification can be also be observed within the three-terminal setup(see appendix C and Fig.~\ref{fig6} for details).

In the following, we try to explore the underlying mechanism of the NDTC in the weak qubit-phonon coupling limit at resonance($\varepsilon=\omega_0$).

% mechanism of NDTC

\subsection{Mechanism of NDTC}
\begin{figure}[tbp]
\begin{center}
%\vspace{-1.0cm}
\includegraphics[scale=0.25]{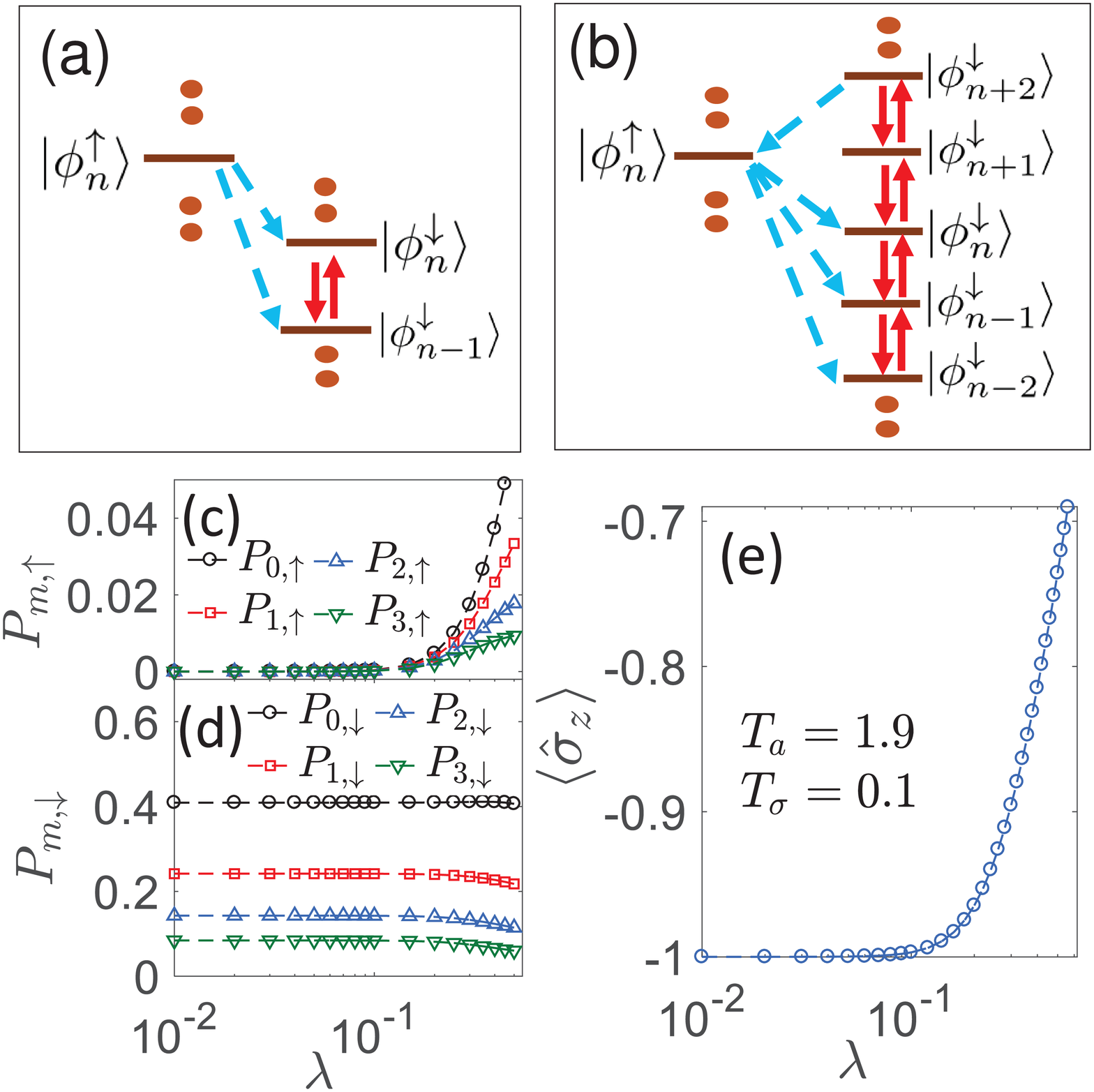}
%\vspace{-2.0cm}
\end{center}
\caption{(Color online)
(a) Schematic illustration of eigenstate transitions in the limiting temperature regime($T_a=1.9$ and $T_\sigma=0.1$) with weak qubit-phonon coupling,
where vertical brown solid lines denote the eigenstate $|\phi^{\eta}_n{\rangle}$  at Eq.~(\ref{state1}) and Eq.~(\ref{state11}),
up and down red solid lines with arrows represent transitions between eigenstates $|\phi^{\eta}_{n}{\rangle}$ and
$|\phi^{\eta}_{n-1}{\rangle}$ at Eq.~(\ref{gap}) and Eq.~(\ref{gam}),
the blue dashed lines with arrows show transitions from $|\phi^{\uparrow}_{n}{\rangle}$ to $|\phi^{\downarrow}_n{\rangle}$($|\phi^{\downarrow}_{n-1}{\rangle}$) at Eq.~(\ref{p1weak}) and Eq.~(\ref{p0weak});
(b) schematic illustration of eigenstate transitions beyond the weak qubit-phonon interaction from Eq.~(\ref{dnmstrong}), where the blue dashed-dotted lines describe
transitions between $|\phi^{\uparrow}_n{\rangle}$ and $|\phi^{\downarrow}_{n{\pm}2}{\rangle}$;
(c) and (d) describe the low-excited steady state populations $|\phi^{\uparrow(\downarrow)}_n{\rangle}$ as a function of $\lambda$;
(e) the steady state qubit bias ${\langle}\hat{\sigma}_z{\rangle}$ by tuning $\lambda$.
The other parameters are given by
$\omega_0=1$, $\varepsilon=1$, $\alpha_a=\alpha_\sigma=0.005$, and $\omega_c=10$.
}~\label{fig4}
\end{figure}

We devote to understanding the NDTC effect with weak qubit-phonon interaction(e.g., $\lambda{<}0.1$).
%In the linear response limit(${\Delta}T{\ll}T_0$), it is expected to see the linear increase of the current(i.e. $J_a{\propto}{\Delta}T$).
To analyze the NDTC, the transition processes at large temperature bias limit(e.g., $T_a{\approx}2$ and $T_\sigma{\approx}0$) are crucial.
The population dynamics $P_{n,\eta}$ with large temperature bias are simplified as
%\begin{eqnarray}
%\frac{dP_{n,1}}{dt}&=&n[\kappa^{+}_a(\omega_0)P_{n-1,1}-\kappa^-_a(\omega_0)P_{n,1}]
%+(1+n)[\kappa^-_a(\omega_0)P_{n+1,1}-\kappa^+_a(\omega_0)P_{n,1}]\\
%&&+[\kappa^+_{\sigma}(\varepsilon)P_{n,0}-\kappa^-_{\sigma}(\varepsilon)P_{n,1}]
%+(1+n)(2\lambda/\omega_0)^2[\kappa^+_{\sigma}(\omega_0+\varepsilon)P_{n-1,0}-\kappa^-_{\sigma}(\omega_0+\varepsilon))P_{n,1}]\nonumber
%\end{eqnarray}
%and
%\begin{eqnarray}
%\frac{dP_{n,0}}{dt}&=&n[\kappa^+_a(\omega_0)P_{n-1,0}-\kappa^-_a(\omega_0)P_{n,0}]
%+(1+n)[\kappa^-_a(\omega_0)P_{n+1,0}-\kappa^+_a(\omega_0)P_{n,0}]\\
%&&-[\kappa^+_{\sigma}(\varepsilon)P_{n,0}-\kappa^-_{\sigma}(\varepsilon)P_{n,1}]
%-(1+n)(2\lambda/\omega_0)^2[\kappa^+_{\sigma}(\omega_0+\varepsilon)P_{n,0}+\kappa^-_{\sigma}(\omega_0+\varepsilon)P_{n+1,1}]\nonumber
%\end{eqnarray}
%\begin{widetext}
\begin{subequations}
\begin{align}
\frac{dP_{n,\uparrow}}{dt}{\approx}&n[\kappa^{+}_a(\omega_0)P_{n-1,\uparrow}-\kappa^-_a(\omega_0)P_{n,\uparrow}]~\label{p1weak}\\
&+(1+n)[\kappa^-_a(\omega_0)P_{n+1,\uparrow}-\kappa^+_a(\omega_0)P_{n,\uparrow}]\nonumber\\
&-[\kappa^-_{\sigma}(\varepsilon)
+(1+n)(2\lambda/\omega_0)^2\kappa^-_{\sigma}(\omega_0+\varepsilon)]P_{n,\uparrow},\nonumber\\
\frac{dP_{n,\downarrow}}{dt}{\approx}&n[\kappa^+_a(\omega_0)P_{n-1,\downarrow}-\kappa^-_a(\omega_0)P_{n,\downarrow}]~\label{p0weak}\\
&+(1+n)[\kappa^-_a(\omega_0)P_{n+1,\downarrow}-\kappa^+_a(\omega_0)P_{n,\downarrow}]\nonumber\\
&+\kappa^-_{\sigma}(\varepsilon)P_{n,\uparrow}
+(1+n)(2\lambda/\omega_0)^2\kappa^-_{\sigma}(\omega_0+\varepsilon)P_{n+1,\uparrow},\nonumber
\end{align}
\end{subequations}
%\end{widetext}
which is illustrated in Fig.~\ref{fig4}(a).
Due to the unidirectional transfer from the dressed state $|\phi_{n,\uparrow}{\rangle}$ to $|\phi_{n(n-1),\downarrow}{\rangle}$,
it is easy to know that populations associated with spin-up are almost fully depleted at steady state, i.e. $P_{n,\uparrow}{\approx}0$ in Fig.~\ref{fig4}(c).
This directly results in the strong localization of the qubit, i.e. ${\langle}\hat{\sigma}_z{\rangle}{\approx}-1$ in Fig.~\ref{fig4}(e),
which significantly blocks the transition from the qubit to the $\sigma$th bath.
While the other branch of populations $P_{n,\downarrow}$ under the thermal equilibrium is distributed as[see Fig.~\ref{fig4}(d)]
\begin{eqnarray}
P^{ss}_{n,\downarrow}{\approx}[1-e^{-\omega/(k_BT_a)}]e^{-n\omega_0/(k_BT_a)}.
\end{eqnarray}
Therefore, the steady state heat current at Eq.~(\ref{ja1}) nearly vanishes($J_a{\approx}0$) in the large temperature bias limit,
which contributes to the emergence of the NDTC.

Moreover, as the qubit-phonon coupling strength is beyond the weak coupling limit(e.g., $\lambda>0.1$), we may expand the coefficient $D_{nm}(2\lambda/\omega_0)$
up to the second order of $\lambda/\omega_0$
\begin{eqnarray}~\label{dnmstrong}
D_{nm}(\frac{2\lambda}{\omega_0})&{\approx}&(-1)^n\{[1-(n+1/2)(\frac{2\lambda}{\omega_0})^2]\delta_{n,m}\nonumber\\
&&+\frac{2\lambda}{\omega_0}(\sqrt{n+1}\delta_{n,m-1}-\sqrt{n}\delta_{n,m+1})\nonumber\\
&&+\frac{1}{2}(\frac{2\lambda}{\omega_0})^2[\sqrt{n(n-1)}\delta_{n,m+2}\nonumber\\
&&+\sqrt{(n+1)(n+2)}\delta_{n,m-2}]\}
\end{eqnarray}
The transition from $|\phi^\downarrow_{n+2}{\rangle}$ to $|\phi^\uparrow_{n}{\rangle}$[see Fig.~\ref{fig4}(b)] is included,
which keeps $P_{n,\uparrow}$ finite, shown in Fig.~\ref{fig4}(c).
Meanwhile, the qubit becomes delocalized[${\langle}\hat{\sigma}_z{\rangle}>-1$ as shown in Fig.~\ref{fig4}(e)].
Hence, it enables the energy exchange between the qubit and the $\sigma$th bath.
Finally, the heat current keeps finite even at large temperature bias limit,
which partially explains the suppression of the NDTC beyond weak qubit-phonon interaction.

%%==========================================
%%==========================================

%%==========================================
\begin{figure}[tbp]
\begin{center}
%\vspace{-1.0cm}
\includegraphics[scale=0.5]{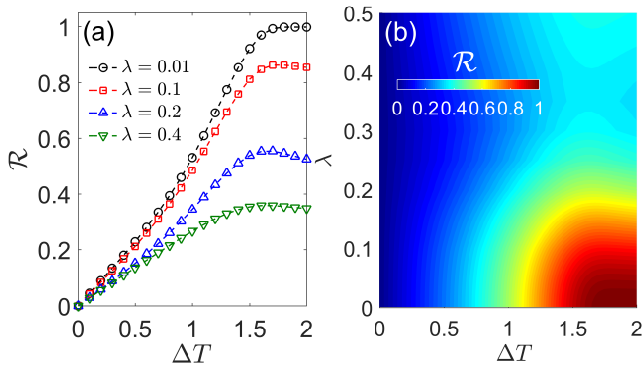}
%\vspace{-2.0cm}
\end{center}
\caption{(Color online) Thermal rectification $\mathcal{R}$ (a) as function of the temperature bias ${\Delta}T$
with $T_a=T_0+{\Delta}T/2$, $T_\sigma=T_0-{\Delta}T/2$ and $T_0=1$,
and (b) both modulating ${\Delta}T$ and the qubit-phonon coupling strength $\lambda$.
The other parameters are $\omega_0=1$, $\varepsilon=1$, $\alpha_a=\alpha_\sigma=0.005$, $\omega_c=10$.
}~\label{fig5}
\end{figure}
%%==========================================
\subsection{Thermal rectification}
Inspired by the asymmetric behavior of the heat current $J_{ss}$ in Fig.~\ref{fig3}(a),
we investigate the thermal rectification effect by tuning temperature bias ${\Delta}T=T_a-T_\sigma$ in Fig.~\ref{fig5}.
The thermal rectification is described as the heat current is larger in one direction than the counterpart in the opposite direction~\cite{bwli2004prl,nbli2012rmp,dsegal2005prl}.
The rectification factor is defined as~\cite{lfzhang2009prb,lfzhang2010prb}
\begin{eqnarray}
\mathcal{R}=\frac{|J_{ss}({\Delta}T)+J_{ss}(-{\Delta}T)|}{\max\{|J_{ss}({\Delta}T)|,|J_{ss}(-{\Delta}T)|\}},
\end{eqnarray}
where $J_{ss}({\Delta}T)$ stands for the current with the temperatures $T_{a}=T_0+{\Delta}T/2$ and $T_{\sigma}=T_0-{\Delta}T/2$.
The thermal rectification becomes most significant as $\mathcal{R}=1$, and it vanishes when $\mathcal{R}=0$.
In Fig.~\ref{fig5}(a), it is shown that temperature bias generally  enhances the rectification factor from weak to strong qubit-phonon couplings.
However, For given ${\Delta}T$ the factor $\mathcal{R}$ is monotonically suppressed by increasing the qubit-phonon coupling strength.
In particularly, the thermal rectification factor approaches unit with weak qubit-phonon interaction(e.g., $\lambda=0.01$)
and large temperature bias limit(${\Delta}T{\approx}2$), and is quite stable.
Such perfect rectification becomes more apparent within the 3D view of $\mathcal{R}$ in Fig.~\ref{fig5}(b).
Therefore, we conclude that the qubit-phonon hybrid system has the great potential to be a perfect thermal rectifier,
and the significant thermal rectification favors weak qubit-phonon coupling.

\section{Conclusion}

To summarize, we investigate quantum heat transfer and multifunctional thermal operations in the nonequilibrium qubit-phonon hybrid system,
which constitutes of one two-level qubit interacting with a single mode phononic field, each weakly coupled to a thermal bath.
We apply the quantum dressed master equation to study dynamics of the qubit-phonon hybrid system by combining with the coherent bosonic state method,
which enables us to study the heat flow with arbitrary qubit-phonon interaction strength.
The effect of the qubit-phonon coupling strength on the steady state heat current is analyzed.
Specifically, it is found that in the weak coupling regime the current is enhanced by the qubit-phonon interaction, which is analytically estimated
as $J_{ss}{\propto}(\lambda/\omega_0)^2$.
On the contrary, in the strong coupling regime the current is significantly suppressed by increasing the coupling strength,
which is mainly due to transition blockade between different eigenstates by dramatic phonon scattering.

Then, we study the effect of the temperature bias $\Delta{T}=T_a-T_\sigma$ on the behavior of the steady state heat current.
In the weak qubit-phonon coupling and large temperature bias regime, it is found that the heat shows astonishing decrease by increasing $\Delta{T}$,
which is a clear signature of the NDTC.
To unravel the underlying mechanism of the NDTC, we analyze the transition process from dynamical equations at Eq.~(\ref{p1weak}) and Eq.~(\ref{p0weak}).
the steady state populations corresponding to the qubit state $|\uparrow{\rangle}$ is almost depleted as $T_\sigma=0$,
which eliminates the energy exchange between the qubit and the $\sigma$th thermal bath.
Moreover, we investigate the influence of the qubit-phonon interaction on the phononic rectification.
The perfect heat rectification($\mathcal{R}=1$) is observed with weak qubit-phonon coupling and large temperature bias,
which corresponds to the significant NDTC.
%Finally, we extend the quantum hybrid system to a single phononic mode coupled to two qubits case to study the heat amplification effect.
%The giant amplification factor is exhibited at moderate gate-temperature regime, which results from the nonmonotonic behavior of the current into the gate-bath.

We hope the analysis of quantum heat transfer and thermal management in the qubit-phonon hybrid system
may have potential applications for the efficient energy control and logical operations of phononic HQSs.
Moreover, we stress that the present work focuses on the energy transfer purely driven by the temperature bias.
The further investigation of influence of the quantum correlation on the energy transport and energy management in the phononic HQSs should be intriguing
to conduct in future~\cite{kmicadei2019nc}.
%%==========================================
%\begin{figure}[tbp]
%\begin{center}
%\vspace{-1.0cm}
%\includegraphics[scale=0.42]{fig3ab.eps}
%\vspace{-2.0cm}
%\end{center}
%\caption{(Color online) Normalized heat current $I/I_{max}(\alpha)$ by
%(a) tuning temperature difference ${\Delta}T=T_L-T_R$ with typical qubit-bath coupling strength $\alpha$
%and (b) both modulating ${\Delta}T$ and $\alpha$ in a 3D view,
%with $I_{\max}(\alpha)=\max_{{\Delta}T}\{I\}$ for a given $\alpha$.
%The temperatures are given by $T_L=T_0+{\Delta}T/2$ and $T_R=T_0-{\Delta}T/2$ with $T_0=2$.
%The other parameters are  $\varepsilon=1$, $\Delta=1$, $U=0.1$, $\omega_c=5$,
%}~\label{fig3}
%\end{figure}
%%==========================================

\section{Acknowledgement}
W.C. would like to thank Jie-Qiao Liao for helpful discussions.
W.C. is supported by the National Natural Science Foundation of China under Grant No. 11704093
and the Opening Project of Shanghai Key Laboratory of Special Artificial Microstructure Materials and Technology.
W.L.Q. and R.J. acknowledge the support by the National Natural Science Foundation of China (No. 11775159), Natural Science Foundation of Shanghai (No. 18ZR1442800), and the National Youth 1000 Talents Program in China.

\appendix

\begin{widetext}
\section{Dynamical equation of populations with weak qubit-phonon coupling}
Following the coefficient at Eq.~(\ref{dnmweak}), the dynamical equations of the system density matrix elements at Eq.~(\ref{dpn}) are specified as
\begin{subequations}
\begin{align}
\frac{dP_{m,\uparrow}}{dt}=&m[\kappa^{+}_a(\omega_0)P_{m-1,\uparrow}-\kappa^-_a(\omega_0)P_{m,\uparrow}]
+(1+m)[\kappa^-_a(\omega_0)P_{m+1,\uparrow}-\kappa^+_a(\omega_0)P_{m,\uparrow}]~\label{popA1}\\
&+[\kappa^+_{\sigma}(\varepsilon)P_{m,\downarrow}-\kappa^-_{\sigma}(\varepsilon)P_{m,\uparrow}]
+m(2\lambda/\omega_0)^2[\kappa^+_{\sigma}(\omega_0+\varepsilon)P_{m-1,\downarrow}-\kappa^-_{\sigma}(\omega_0+\varepsilon))P_{m,\uparrow}]\nonumber\\
&+(1+m)(2\lambda/\omega_0)^2[\kappa^-_{\sigma}(\omega_0-\varepsilon)P_{m+1,\downarrow}-\kappa^+_\sigma(\omega_0-\varepsilon)P_{m,\uparrow}]\nonumber\\
&+(1+m)(2\lambda/\omega_0)^2[\kappa^+_{\sigma}(\varepsilon-\omega_0)P_{m+1,\downarrow}-\kappa^-_\sigma(\varepsilon-\omega_0)P_{m,\uparrow}],\nonumber\\
\frac{dP_{m,\downarrow}}{dt}=&m[\kappa^+_a(\omega_0)P_{m-1,\downarrow}-\kappa^-_a(\omega_0)P_{m,\downarrow}]
+(1+m)[\kappa^-_a(\omega_0)P_{m+1,\downarrow}-\kappa^+_a(\omega_0)P_{m,\downarrow}]~\label{popA2}\\
&-[\kappa^+_{\sigma}(\varepsilon)P_{m,\downarrow}-\kappa^-_{\sigma}(\varepsilon)P_{m,\uparrow}]
-(1+m)(2\lambda/\omega_0)^2[\kappa^+_{\sigma}(\omega_0+\varepsilon)P_{m,\downarrow}-\kappa^-_{\sigma}(\omega_0+\varepsilon)P_{m+1,\uparrow}]\nonumber\\
&-m(2\lambda/\omega_0)^2[\kappa^-_{\sigma}(\omega_0-\varepsilon)P_{m,\downarrow}-\kappa^+_\sigma(\omega_0-\varepsilon)P_{m-1,\uparrow}]\nonumber\\
&-m(2\lambda/\omega_0)^2[\kappa^+_{\sigma}(\varepsilon-\omega_0)P_{m,\downarrow}-\kappa^-_\sigma(\varepsilon-\omega_0)P_{m-1,\uparrow}],\nonumber
\end{align}
\end{subequations}
where the rates are $\kappa^{+}_{u}(\omega)=\theta(\omega)\gamma_u(\omega)n_u(\omega)$
and $\kappa^{-}_{u}(\omega)=\theta(\omega)\gamma_u(\omega)[1+n_u(\omega)]$,
with $\theta(\omega)=1$ for $\omega>0$ and $\theta(\omega)=0$ for $\omega{\leq}0$.

%%==========================================
\begin{figure}[tbp]
\begin{center}
%\vspace{-1.0cm}
\includegraphics[scale=0.3]{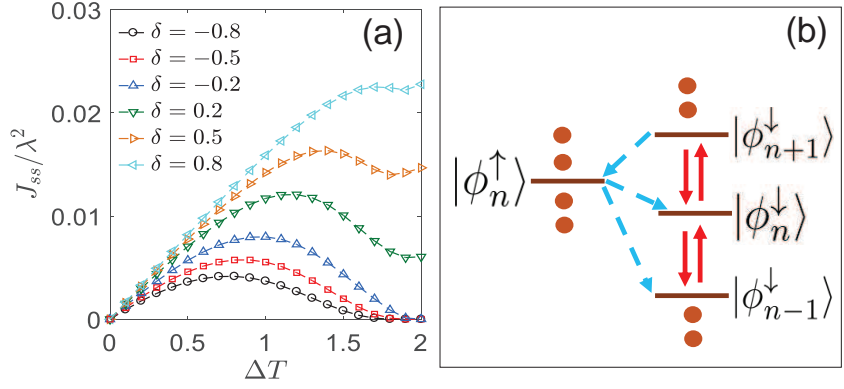}
%\vspace{-2.0cm}
\end{center}
\caption{(Color online)
(a)Effect of the qubit bias $\delta=\omega_0-\varepsilon$ on the steady state heat current $J_{ss}/\lambda^2$ as a function of the temperature bias ${\Delta}T$, with $T_a=T_0+{\Delta}T/2$, $T_\sigma=T_0-{\Delta}T/2$, and $T_0=1$;
(b) the transitions at Eq.~(\ref{gpm-weak})(blue dashed line with arrow), Eq.~(\ref{gap}) and Eq.~(\ref{gam})(red solid line with arrow) at off-resonance($\delta{>}0$) in the limiting regime(${\Delta}T{\approx}2$).
The other parameters are $\omega_0=1$, $\lambda=0.01$, $\alpha_a=\alpha_\sigma=0.005$, and $\omega_c=10$.
}~\label{fig8}
\end{figure}
%%==========================================

\section{Effect of the energy bias $(\delta=\omega_0{-}\varepsilon)$ on the NDTC}
We analyze the effect of the energy bias $\delta$ on the renormalized steady state heat current $J_{ss}/\lambda^2$ in Fig.~\ref{fig8}(a).
It is found that it is found that by increasing the detuning $\delta$, the current shows monotonic enhancement.
While for the NDTC, in the regime $\delta<0$ it always becomes significant.
While in the regime $\delta>0$, the signature of the NDTC gradually becomes suppressed with the increase of $\delta$.
In particular, the NDTC completely vanishes at large detuning limit(e.g., $\varepsilon=0.2\omega_0$).
In the limiting temperature regime(e.g., $T_a{\approx}2$ and $T_{\sigma}{\approx}0$),
from Eq.~(\ref{popA1}) and Eq.~(\ref{popA2}) it is known that besides transitions
 $|\phi^{\uparrow}_{n}{\rangle}{\rightarrow}|\phi^{\downarrow}_{n}{\rangle}$
 and
  $|\phi^{\uparrow}_{n}{\rangle}{\rightarrow}|\phi^{\downarrow}_{n-1}{\rangle}$,
there exists the additional transition from $|\phi^{\downarrow}_{n+1}{\rangle}$ to $|\phi^{\uparrow}_{n}{\rangle}$ in Fig.~(\ref{fig8})(b).
Hence, it avoids the populations $P_{n,\uparrow}$ from depletion,
which mainly results in the finite heat current(e.g., blue dashed line with left-triangles).
This partially explains the suppression of the NDTC.

\section{Dynamical equation of two qubits coupled to one bosonic field}

\subsection{Model}
The Hamiltonian of two qubits coupled to a cavity is described as
\begin{eqnarray}
\hat{H}_s=\frac{1}{2}\sum_{v=L,R}\varepsilon_v\hat{\sigma}^v_z+\omega_0\hat{a}^{\dag}\hat{a}+\sum_{v=L,R}\lambda^\sigma_v\hat{\sigma}^v_z(\hat{a}^{\dag}+\hat{a}),
\end{eqnarray}
Under the spin basis $\{|1{\rangle}=|\uparrow\uparrow{\rangle},|2{\rangle}=|\uparrow\downarrow{\rangle},
|3{\rangle}=|\downarrow\uparrow{\rangle},|4{\rangle}=|\downarrow\downarrow{\rangle}\}$, the system Hamiltonian can be exactly solved as
\begin{subequations}
\begin{align}
\hat{H}_s|1{\rangle}=&[\omega_0(\hat{a}^{\dag}+\lambda/\omega_0)(\hat{a}+\lambda/\omega_0)-\lambda^2/\omega+\overline{\varepsilon}/2]|1{\rangle},\\
\hat{H}_s|2{\rangle}=&[\omega_0(\hat{a}^{\dag}+{\delta}\lambda/\omega_0)(\hat{a}+{\delta}\lambda/\omega_0)-({\delta}\lambda)^2/\omega_0+{\delta}\varepsilon/2]|2{\rangle},\\
\hat{H}_s|3{\rangle}=&[\omega_0(\hat{a}^{\dag}-{\delta}\lambda/\omega_0)(\hat{a}-{\delta}\lambda/\omega_0)-({\delta}\lambda)^2/\omega_0-{\delta}\varepsilon/2]|3{\rangle},\\
\hat{H}_s|4{\rangle}=&[\omega_0(\hat{a}^{\dag}-\lambda/\omega_0)(\hat{a}-\lambda/\omega_0)-\lambda^2/\omega_0-\overline{\varepsilon}/2]|4{\rangle},
\end{align}
\end{subequations}
with $\lambda=\lambda^\sigma_L+\lambda^\sigma_R$, ${\delta}\lambda=\lambda^\sigma_L-\lambda^\sigma_R$,
$\overline{\varepsilon}=\varepsilon_1+\varepsilon_2$
and ${\delta}\varepsilon=\varepsilon_1-\varepsilon_2$.
Hence, for the spin state $|\eta{\rangle}$, we introduce the displaced photon state
$|n{\rangle}_{\eta}=\frac{(\hat{a}^{\dag}+g_\eta)^{n}}{\sqrt{n!}}\exp(-g^2_\eta/2+g_\eta\hat{a}^{\dag})|0{\rangle}_p$
with the bare vacuum state $\hat{a}|0{\rangle}_p=0$.
$|n,\eta{\rangle}=|n{\rangle}_{\eta}{\otimes}|\eta{\rangle}$ constitutes the eigenstate of $\hat{H}_s$
with the eigenvalue $E^\eta_{n}=\omega_0{n}+\Lambda_\eta$,
where the displaced coefficients are
$g_1=(\lambda^\sigma_L+\lambda^\sigma_R)/\omega_0$, $g_2=(\lambda^\sigma_L-\lambda^\sigma_R)/\omega_0$,
$g_3=(-\lambda^\sigma_L+\lambda^\sigma_R)/\omega_0$, $g_4=(-\lambda^\sigma_L-\lambda^\sigma_R)/\omega_0$,
and the displaced energies are
$\Lambda_1=(\varepsilon_L+\varepsilon_R)/2-(\lambda^\sigma_L+\lambda^\sigma_R)^2/\omega_0$,
$\Lambda_2=(\varepsilon_L-\varepsilon_R)/2-(\lambda^\sigma_L-\lambda^\sigma_R)^2/\omega_0$,
$\Lambda_3=(-\varepsilon_L+\varepsilon_R)/2-(\lambda^\sigma_L-\lambda^\sigma_R)^2/\omega_0$
and
$\Lambda_4=(-\varepsilon_L-\varepsilon_R)/2-(\lambda^\sigma_L+\lambda^\sigma_R)^2/\omega_0$.

Three thermal baths are given by
$\hat{H}_b=\sum_{v=L,M,R}\omega_k\hat{b}^{\dag}_{k,v}\hat{b}_{k,v}$.
The system-bath interaction is given by
\begin{eqnarray}
\hat{H}_{sb}=\sum_{v,k}(f_{k,v}\hat{b}^{\dag}_{k,v}\hat{S}_{v}+f^{*}_{k,v}\hat{S}^{\dag}_{v}\hat{b}_{k,v}),
\end{eqnarray}
with $\hat{S}_{L(R)}=\hat{\sigma}^{L(R)}_{x}$ and $\hat{S}_{M}=\hat{a}$.

\subsection{Quantum master equation}
Under the eigenbasis $\{|n,\eta{\rangle}\}~\eta=1,2,3,4$, quantum master equation is given by
\begin{eqnarray}
\frac{d\hat{\rho}_{s}(t)}{dt}&=&\sum_{v;n,m;\eta,\eta^{\prime}}\Theta(\Delta^{m,\eta^{\prime}}_{n,\eta})\gamma_v(\Delta^{m,\eta^{\prime}}_{n,\eta})|{\langle}n,\eta|\hat{S}^{\dag}_v|m,\eta^{\prime}{\rangle}|^2\{n_v(\Delta^{m,\eta^{\prime}}_{n,\eta})
\mathcal{D}[|n,\eta{\rangle}{\langle}m,\eta^{\prime}|]\hat{\rho}_{s}(t)\nonumber\\
&&+[1+n_v(\Delta^{m,\eta^{\prime}}_{n,\eta})]\mathcal{D}[|m,\eta^{\prime}{\rangle}{\langle}n,\eta|]\hat{\rho}_{s}(t)\},
\end{eqnarray}
where the dissipator is
\begin{eqnarray}
\mathcal{D}[|m,\eta^{\prime}{\rangle}{\langle}n,\eta|]\hat{\rho}_s=
|m,\eta^{\prime}{\rangle}{\langle}n,\eta|\hat{\rho}_s|n,\eta{\rangle}{\langle}m,\eta^{\prime}|
-\frac{1}{2}|n,\eta{\rangle}{\langle}n,\eta|\hat{\rho}_s
-\frac{1}{2}\hat{\rho}_s|n,\eta{\rangle}{\langle}n,\eta|,
\end{eqnarray}
with the energy gap
$\Delta^{m,\eta^{\prime}}_{n,\eta}=E_{n,\eta}-E_{m,\eta^{\prime}}$,
 and the transition coefficients
\begin{subequations}
\begin{align}
{\langle}n,\eta|\hat{\sigma}^{L,R}_x|m,\eta^{\prime}{\rangle}=&{_\eta}{\langle}n|m{\rangle}_{\eta^{\prime}}
{\langle}\eta|\hat{\sigma}^{L,R}_x|\eta^{\prime}{\rangle},\\
{\langle}n,\eta|\hat{a}^{\dag}|m,\eta^{\prime}{\rangle}=&{_\eta}{\langle}n|\hat{a}^{\dag}|m{\rangle}_{\eta}.
\end{align}
\end{subequations}
Specifically,
\begin{subequations}
\begin{align}
{\langle}n,1|\hat{\sigma}^L_x|m,3{\rangle}=&{\langle}n,2|\hat{\sigma}^L_x|m,4{\rangle}=(-1)^nD_{nm}(2\lambda^\sigma_L/\omega),\\
{\langle}n,1|\hat{\sigma}^R_x|m,2{\rangle}=&{\langle}n,3|\hat{\sigma}^R_x|m,4{\rangle}=(-1)^nD_{nm}(2\lambda^\sigma_R/\omega),\\
{\langle}n,\eta|\hat{a}^\dag|m,\eta{\rangle}=&\sqrt{m+1}\delta_{n,m+1}-g_\eta\delta_{n,m}.
\end{align}
\end{subequations}
Then, the dynamical equation of the elements is given by
\begin{eqnarray}
\frac{dP_{n,\eta}}{dt}&=&\sum_{v;n^{\prime},\eta^\prime}
[\Gamma^{v,+}_{(n^\prime,\eta^\prime){\rightarrow}(n,\eta)}P_{n^{\prime},\eta^\prime}
-\Gamma^{v,+}_{(n,\eta){\rightarrow}(n^\prime,\eta^\prime)}P_{n,\eta}]\nonumber\\
&&+\sum_{v;n^{\prime},\eta^\prime}
[\Gamma^{v,-}_{(n^\prime,\eta^\prime){\rightarrow}(n,\eta)}P_{n^{\prime},\eta^\prime}
-\Gamma^{v,-}_{(n,\eta){\rightarrow}(n^\prime,\eta^\prime)}P_{n,\eta}],
\end{eqnarray}
where the transition rates are
\begin{subequations}
\begin{align}
\Gamma^{v,+}_{(n^\prime,\eta^\prime){\rightarrow}(n,\eta)}=&\Theta(\Delta^{n^\prime,\eta^\prime}_{n,\eta})\gamma_v(\Delta^{n^\prime,\eta^\prime}_{n,\eta})n_v(\Delta^{n^\prime,\eta^\prime}_{n,\eta})
|{\langle}n,\eta|\hat{S}^{\dag}_v|n^{\prime},\eta^{\prime}{\rangle}|^2\\
\Gamma^{v,-}_{(n^\prime,\eta^\prime){\rightarrow}(n,\eta)}=&\Theta(\Delta^{n,\eta}_{n^\prime,\eta^\prime})\gamma_v(\Delta^{n,\eta}_{n^\prime,\eta^\prime})[1+n_v(\Delta^{n,\eta}_{n^\prime,\eta^\prime})]
|{\langle}n^{\prime},\eta^{\prime}|\hat{S}^{\dag}_v|n,\eta{\rangle}|^2,
\end{align}
\end{subequations}
with $\Theta(x)=1$ for $x>0$, and $\Theta(x)=0$ for $x<=0$,
the spectral function of the $v$th thermal bath
$\gamma_v(\omega)=\alpha_v\omega\exp(-|\omega|/\omega_{c,v})$
and the Bose-Einstein distribution function
$n_v(\omega)=1/[\exp(\omega/k_BT_v)-1]$.
Accordingly,  the steady state current into the $\sigma_v$th thermal bath is expressed as
\begin{eqnarray}~\label{current2}
J^{\sigma}_v=\sum_{n,m;\eta,\eta^\prime}[\Delta^{m,\eta^\prime}_{n,\eta}
\Gamma^{v,-}_{(n,\eta){\rightarrow}(m,\eta^\prime)}P^{ss}_{n,{\eta}}
-\Delta^{n,\eta}_{m,\eta^\prime}
\Gamma^{v,+}_{{(n,\eta){\rightarrow}(m,\eta^\prime)}}
P^{ss}_{n,{\eta}}],
\end{eqnarray}
and the current into the $a$th bath is given by
\begin{eqnarray}~\label{current3}
J_a=\sum_{n,m;\eta,\eta^\prime}[\Delta^{m,\eta^\prime}_{n,\eta}
\Gamma^{a,-}_{(n,\eta){\rightarrow}(m,\eta^\prime)}P^{ss}_{n,{\eta}}
-\Delta^{n,\eta}_{m,\eta^\prime}
\Gamma^{a,+}_{{(n,\eta){\rightarrow}(m,\eta^\prime)}}
P^{ss}_{n,{\eta}}],
\end{eqnarray}
where the energy gap is $\Delta^{m,\eta^\prime}_{n,\eta}=E^{\eta^\prime}_{m}-E^{\eta}_{n}$.
% the expression of heat currents

%%==========================================
\begin{figure}[tbp]
\begin{center}
%\vspace{-1.0cm}
\includegraphics[scale=0.35]{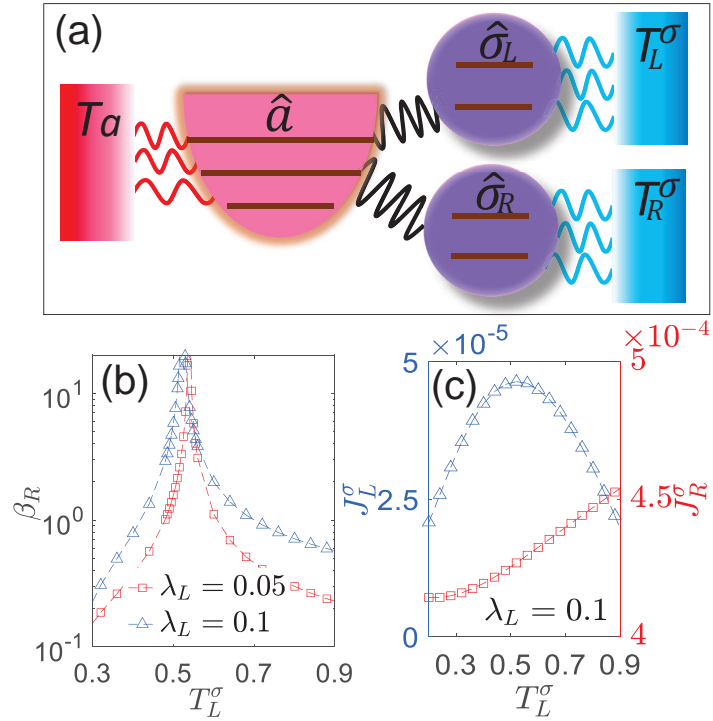}
%\vspace{-2.0cm}
\end{center}
\caption{(Color online) (a) Schematic illustration of  one phononic mode(pink half-circle marked as  $\hat{a}$)
coupled to two qubits(blue circles marked as $\hat{\sigma}_{L(R)}$),
each individually interacting with a thermal bath characterized as temperatures $T_{a}$, $T^\sigma_L$, and $T^\sigma_R$, respectively;
(b) heat amplification factor $\beta_R$ at Eq.~(\ref{betar}) with left-qubit phonon coupling strength $\lambda^\sigma_L=0.05, 0.1$,
and (c) heat currents $J^{\sigma}_L$ and $J^\sigma_R$ at Eq.~(\ref{current2}) with $\lambda^\sigma_L=0.1$ as a function of $T^\sigma_L$,
both with right-qubit phonon coupling strength given by $\lambda^\sigma_R=4\lambda^\sigma_L$.
The other parameters are $\omega_0=1$, $\varepsilon=1$, $\alpha_a=\alpha_\sigma=0.005$, $\omega_c=10$,
$T_a=1.2$, and $T^\sigma_R=0.2$.
}~\label{fig6}
\end{figure}
%%==========================================
\subsection{Heat amplification}

To analyze the heat amplification effect, we set the $a$th bath as the hot source, the $\sigma_R$th bath as the cold drain,
and the $\sigma_L$th bath as the gate with the tunable temperature $T^\sigma_L{\in}[T^\sigma_R,T_a]$.
The amplification factor is defined as~\cite{nbli2012rmp}
\begin{eqnarray}~\label{betar}
\beta_R=\Big{|}\frac{{\partial}J^\sigma_R}{{\partial}J^\sigma_L}\Big{|}.
\end{eqnarray}
The heat amplification occurs once the tiny change of $J^\sigma_L$ may dramatically modulate $J^\sigma_R$,
specified as $\beta_R>1$.
We focus on the heat amplification in the weak qubit-phonon coupling regime in Fig.~\ref{fig6}(b),
in which the NDTC generally appears as shown in Fig.~\ref{fig3}.
It is found that there exists a giant amplification factor in the moderate temperature regime(e.g., $T^\sigma_L{\approx}0.53$ when $\lambda^\sigma_L=0.1$).
Accordingly, the heat current $J^\sigma_L$ is much smaller than $J^\sigma_L$[see Fig.~\ref{fig6}(c)],
which ensures the valid application of this setup as a quantum thermal transistor.
Moreover, it becomes suppressed the temperature $T^\sigma_L$ is tuned away from this giant factor regime.
And it fails to realize the heat amplification in the small and large temperature limits of $T^\sigma_L$.
Though not shown in this paper, it should be noted that other setups can also realize the thermal transistor effect, besides the one shown in Fig.~\ref{fig6}(a),
e.g., exchanging the position of the qubit $\sigma_R$ with the phonon mode $a$.

\end{widetext}

\end{document}